\documentclass[adp,a4paper,fleqn%
 ,earlyview
]{w-art}
\usepackage{times,cite,w-thm}
\theoremstyle{plain}

\theoremstyle{definition}

\usepackage[]{graphicx}
\usepackage{subfigure}
\usepackage{amssymb}
\begin{document}
\renewcommand{\copyrightyear}{2011}
\DOIsuffix{theDOIsuffix}
\Volume{XX}
\Month{XX}
\Year{2012}
\pagespan{1}{}
\Receiveddate{XXXX}
\Reviseddate{XXXX}
\Accepteddate{XXXX}
\Dateposted{XXXX}



\title[2CK]{Rotational Quantum Impurities in a Metal: Stability of the 
2-Channel Kondo Fixed Point in a Magnetic Field}


\author[K. Ballmann]{Katinka Ballmann\inst{}%
}
\address{Physikalisches Institut, Universit\"at Bonn, Nussallee 12,
53115 Bonn, Germany}
\author[J. Kroha]{Johann Kroha\inst{}
\footnote{Corresponding author\quad
E-mail:~\textsf{kroha@physik.uni-bonn.de}
}}
\begin{abstract}
A three-level system with partially broken SU(3) symmetry immersed in a metal, 
comprised of a unique non-interacting ground state and two-fold 
degenerate excited states, exhibits a stable two-channel Kondo fixed point
within a wide range of parameters, as has been shown in previous work. 
Such systems can, for instance, be realized by protons dissolved in 
a metal and bound in the interstitial space of the host lattice, where 
the degeneracy of excited rotational states is guaranteed by the  
space inversion symmetry of the lattice. 
We analyze the robustness of the 2CK fixed point with respect to
a level splitting of the excited states and discuss how this may explain 
the behavior of the well-known dI/dV spectra measured by Ralph and 
Buhrman on ultrasmall quantum point contacts in a magnetic field.  
\end{abstract}
\maketitle                   



\renewcommand{\leftmark}
{K. Ballmann and J. Kroha: Rotational quantum impurities in a metal: 
stability of the 2CK fixed point}



\section{Introduction}
New, exotic quantum states of matter can arise in electronic systems, 
when two degenerate ground states compete with each other and get 
entangled, leading to nontrivial behavior of the entropy and of the 
thermodynamic, magnetic and electric response. 
In the case of the 
spin-1/2 two-channel Kondo (2CK) effect, the spin-screening of a 
spin-1/2 impurity by the first and by the second of two identical, 
conserved conduction electron continua (channels), respectively, 
are in competition for each of these screening channels to form a 
spin singlet \cite{nozieres80}. However, in this system both the 
weak coupling fixed point (decoupled spin-1/2 impurity) and the 
strong-coupling fixed point (a three-body bound state comprised of 
the spin-1/2 impurity and two conduction electrons) are doubly 
degenerate and, hence, are unstable with respect to a small coupling 
to the conduction band \cite{nozieres80,kolf07}. 
As a result, a stable intermediate coupling fixed point is formed,
where an intricate, quantum-frustrated many-body ground state with 
a non-vanishing zero-point entropy of $S(T=0)=k_B \ln \sqrt{2}$
is realized, $k_B$ denoting the Boltzmann constant \cite{andrei84,tsvelik94}.
The spin degree of freedom need not be magnetic spin, but may be any 
two-dimensional representation of SU(2), i.e. a pseudospin 1/2.

Signatures of the 2CK effect have been observed experimentally in 
certain heavy-fermion compounds \cite{seaman91,cichorek05}, where 
the Kondo degree of freedom might arise from orbital degeneracy 
\cite{cox88a,cox88b}. However, not all measured response 
quantities have been found to be in accordance with 2CK behavior in
these systems. More recently, the 2CK fixed point has been 
predicted to exist \cite{oreg03} and then realized \cite{goldhaber07} 
in an an ingeniously designed, fine-tuned semiconductor Qdot system. 
However, perhaps the most intriguing as well as controversial experiments 
regarding the 2CK effect remain the $dI/dV$ spectroscopy measurements 
by Ralph and Buhrman on ultrasmall Cu quantum point contacts 
\cite{ralph92,ralph95}. They exhibit scaling behavior of the 
conductance near zero bias, as expected from a 2CK system 
\cite{affleck91,affleck93,vdelft95}, without fine-tuning of parameters as well 
as sharp conductance spikes at elevated bias. 
Alternative scattering mechanisms \cite{altshuler79,kozub97}, 
different from 2CK physics, cannot account for the complete body of 
experimental observations \cite{hettler94,vondelft98}. A review of 
theoretical and experimental aspects of 2CK physics can be found in 
Ref.~\cite{coxzawa98}.

It is difficult to design realistic, microscopic models which 
generically, i.e., without fine-tuning, exhibit a 2CK fixed point, because 
either the Kondo (pseudo)spin symmetry or the channel degeneracy are 
easily broken. Two-level systems (TLS), e.g., an ion in a double-well 
potential, embedded in a metal, have been put forward early-on as 2CK systems 
and have been intensively studied by Zawadowski and co-workers
\cite{vladar83,vladar88,zawa94}. 
These studies have lead to a profound understanding of the dynamics of TLS 
in metals. However, it turned out that in the standard TLS model of a 
particle in a double-well potential the 2CK regime cannot be realized.
This is because the Kondo coupling and the tunneling rate and, hence, 
the ground state level splitting are not independent parameters in this model.
It was shown that the two-channel Kondo temperature $T_K$ is always smaller 
than the level splitting, because of screening effects \cite{aleiner02} 
and coupling to higher excited states \cite{aleiner01},
so that the physics is always dominated by the level splitting. 
Despite extensions of the TLS model \cite{zarand05} it has remained difficult 
to stabilize a 2CK fixed point.

In order to provide an explanation for the 2CK physics and at the 
same time for the conductance spikes observed in the Ralph-Buhrman
experiments without fine-tuning, we have earlier proposed and analyzed 
the model of a dynamical impurity with a {\it rotational} degree of freedom, 
immersed in a metal \cite{arnold07}. In this rotational impurity model (RIM)
the Kondo SU(2) symmetry is stabilized by the space inversion symmetry of 
the host material, while the channel degree of freedom is the (magnetic) 
conduction electron spin, and its degeneracy is guaranteed by time reversal 
symmetry. Taking the first rotational doublet into account, it was 
shown by perturbative renormalization group (RG) that this model generically
has a stable 2CK fixed point within a wide range of parameter values.
In addition, the conductance spikes were naturally explained within the
same model as Kondo-like transitions between a rotational doublet and
the impurity ground state. In the present paper we extend this study to
the behavior in a magnetic field, Zeeman-coupled to the magnetic moment of
a rotating, charged particle. 
In Section 2 we describe the model and its renormalization group 
treatment in more detail and discuss 
briefly, why charge screening effects or higher excitations of the 
rotational impurity will not suppress $T_K$, in contrast to the case of a  
double-well impurity \cite{aleiner02,aleiner01}. In Section 3 we present 
the results of the perturbative RG for the RG flow of the energy and the 
decay rate of the excited rotational doublet and, in particular, the 
phase diagram of the model in the presence of a magnetic field,
lifting the doublet degeneracy.   
We conclude in Section 4 with a discussion of the implication of the 
results for the interpretation of the conductance spectroscopy measurements
\cite{ralph92,ralph95}.

\section{Microscopic model and renormalization group treatment}
\subsection{Hamiltonian}
\begin{figure}[t]
\centering
\includegraphics[height=3cm]{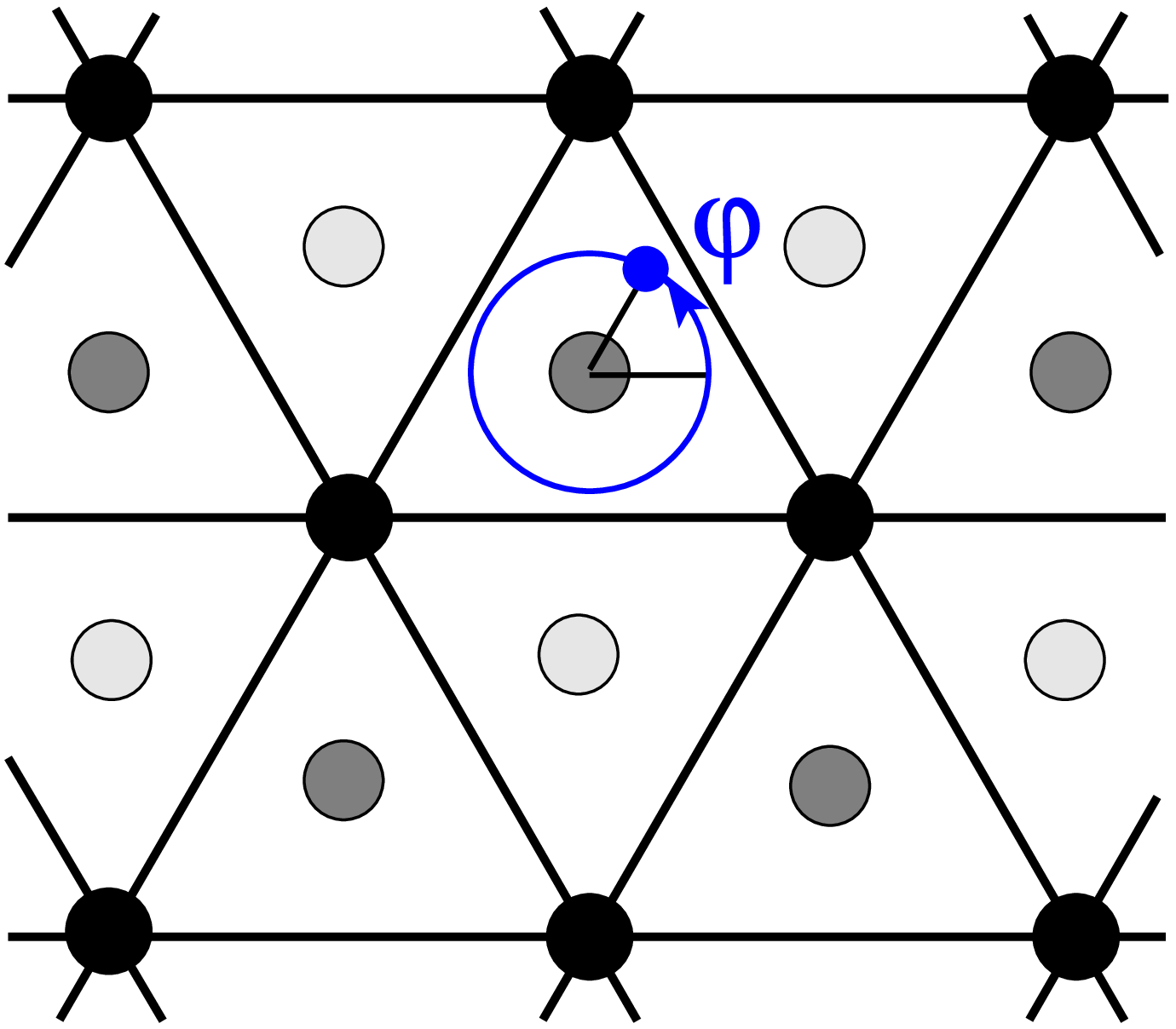}~a)
\hfil
\includegraphics[height=3cm]{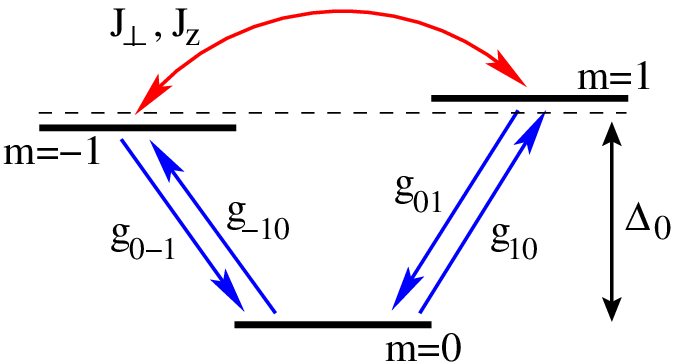}~b)
\hfil
\includegraphics[height=3cm]{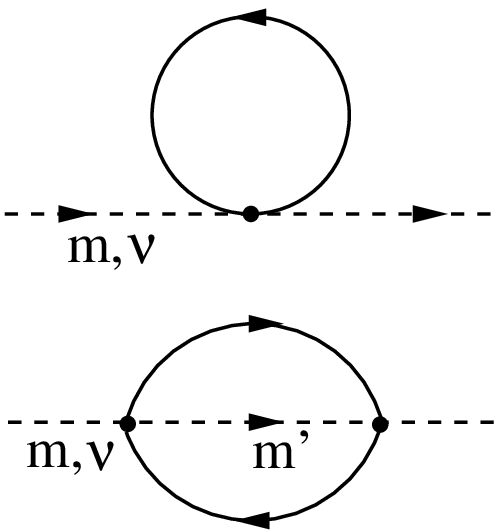}~c)
\caption{\textbf{a)} Cu lattice with a rotational impurity at interstitial 
site. Black dots define one layer of the Cu (111) plane, light and dark
grey dots define the layers above and below, respectively. The blue dot 
is the impurity with a two-fold degenerate rotational degree of freedom.
\textbf{b)} Level scheme of the three-state quantum impurity, 
defining the coupling constants. The additional superscript carried by the 
$g_{mn}^{(j)}$ in Eq.~\ref{eq:hamilton} indicates the initial conduction
electron state before scattering, which is crucial for the RG treatment
of excited states \cite{arnold07}.
\textbf{c)} Impurity self-energy diagrams, where solid lines represent
conduction electron propagators and dashed lines pseudo fermion
progagators.}
\label{fig:model}
\end{figure}

Hydrogen is easily dissolved in ionic form in noble metals like paladium 
or copper. The protons occupy the interstitial spaces of the host lattice. 
If this lattice obeys space inversion symmetry, like the Cu fcc lattice,
all excited states of a proton in the lattice potential are doubly degenerate, 
while it's ground state is unique. The excited-state doublets may be 
visualized as roton states with opposite rotational orientation,
see Fig.~\ref{fig:model} a). Taking only the first excited doublet and the 
ground state into account and considering that conduction electron scattering
can induce transitions between any of these three states, one obtains
a three-level model whose SU(3) symmetry is partially broken, due to the bare 
level splitting $\Delta_0$ between the ground and the excited states
\cite{arnold07}. The level scheme is shown in Fig.~\ref{fig:model} b), 
along with the various couplings of conduction electrons to the
dynamical impurity. The corresponding Hamiltonian reads,
\begin{eqnarray}
H&=&{\sum_{{\bf{k}}\sigma m}}'
\varepsilon_{\mathbf{k}}c_{\mathbf{k}\sigma
m}^\dagger c_{\mathbf{k}\sigma
m}+\Delta_0\sum_{m=\pm1}f_m^\dagger
f_m +B_0\sum_{m=\pm1}m\,f_m^\dagger f_m\nonumber\\
&&+\sum_{\sigma}\Big(\Big[\frac{1}{2}\sum_{i,j=\pm1}ijJ_z^{ij}S_{i,i}{s_
{ j ,j}^\sigma} \Big]+\sum_{i,j=\pm1\atop i\neq
j}J_{\perp}^{i}S_{i,j}{s_{j,i}^\sigma}\Big)\nonumber\\
&&+\sum_{\sigma}\Big(\sum_{m,n\atop-1\leq
n-m\leq1}\Big[g_{m0}^{(n)}S_{m,0}{s_{n-m,n}^\sigma}+H.c.\Big]
+\sum_{m=\pm1}2g_{mm}^{(0)}S_{m,m}{s_{0,0}^\sigma}\Big)\ .
\label{eq:hamilton}
\end{eqnarray} 
The first term is the usual conduction electron kinetic energy. The
conduction electron operators $c_{\mathbf{k}\sigma m}^\dagger$ carry
the {\it conserved} magnetic spin $\sigma=\pm\frac{1}{2}$, acting as the
channel degree of freedom, and additionally an SU(3) index $m=0,\pm1$,
labelling lattice angular momentum states of the conduction electrons 
which may be changed in a scattering process. 
Since $m$ is an orbital degree of freedom, the sum over the continuum of
single-electron states ${\bf k}$ must be restricted such that the double
sum $\sum_{{\bf k} m}'$ covers k-space completely without oversummation. This
restriction is denoted by a prime in the first term of Eq. (1).
The second term describes the two-fold degenerate impurity states,
$m=\pm1$, with bare excitation energy $\Delta_0$ above the impurity ground 
state, $m=0$.  In the presence of a magnetic field coupling to the 
rotational motion (magnetic moment) of the charged impurity particle, 
the excited states
acquire an additional Zeeman splitting, $2B_0$, described by the third term 
in Eq.~(\ref{eq:hamilton}). We use Abrikosov's pseudo-fermion representation 
to describe the impurity levels, where $f_m^\dagger$ is the creation operator 
for the impurity in state $m=0,\pm1$. The defect dynamics are restricted 
by the constraint $Q=\sum_{m=0,\pm1}f_m^\dagger f_m=1$. 
The last two terms represent transitions between the local level induced 
by conduction electron scattering, including the potential scattering term for 
completeness, although it is an irrelevant operator. 
See Fig.~\ref{fig:model} b) for the definition of the coupling constants. 
The impurity operators are defined as $S_{m,n}=f^\dagger_m f_n$, and
the operators acting on the electronic Fock space are obtained by substituting
$f_m\to\sum_{\bf{k}}c_{\bf{k}\sigma m}$ in the above expressions. 

\subsection{Perturbative RG Analysis}
We employ the one-loop (second-order perturbation theory) renormalization 
group to analyze the three-level model, Eq.~(\ref{eq:hamilton}).
Transitions between the non-degenerate impurity states 
($\Delta m=\pm 1$) necessarily involve resonant, inelastic electron 
scattering, where an electron initiates from or ends up in an excited state. 
Therefore, it is not sufficient to calculate renormalized couplings at the 
Fermi energy, but it is crucial to take the dependence of the renormalized 
coupling constants on the (initial) energy of the scattering electrons into 
account \cite{arnold07}. The corresponding formalism has originally been  
developed for non-equlilibrium Kondo systems at finite bias
\cite{rosch03,paaske04a,rosch05}, 
but it is suitable for the present situation as 
well. We take the impurity dynamics on-shell, 
i.e. the pseudofermion energy $\nu$ equal to the respective impurity level 
energy, $\nu=|m|\Delta (D)$, in all expressions. 
The one-loop RG equations for the running couplings $g_{mn}^{(j)}$ then read,
\begin{eqnarray}
\hspace{-0.5cm}\frac{d g_{mn}^{(j)}(\omega)}{d \ln D} &=&
2 \hspace{-0.5cm}\sum _{\stackrel{j \ell }{-1\leq j+n-\ell \leq
1}}\hspace{-0.5cm}g_{m\ell}^{(j+n-\ell)}
   \bigl(\Omega _{n\ell} \bigr) \, 
   g_{\ell n}^{(j)}
   \bigl(\omega \bigr) \, 
   \Theta\Bigl(D-\sqrt{\Omega_{n\ell}^2+\Gamma_{\ell}^2}\Bigr)
   \bigl(1-\delta_{m\ell}\delta_{n\ell}\bigr) - {\rm (exch.)}  
\label{eq:RG}
\end{eqnarray}
with $g_{1,-1}^{(1)}=g_{-1,1}^{(-1)}=J_{\perp}$ 
and $g_{mm}^{(j)} = jm\,J_{z}/2$, for $j=\pm 1$, $m=\pm 1$.
The exchange terms (exch.) on the right-hand side (RHS) of Eq. (\ref{eq:RG}) 
are obtained from the direct ones by interchanging in the $g$'s 
in- and out-going pseudofermion indices and by interchanging 
$m\leftrightarrow n$ everywhere. The energy arguments of the $g$'s on the 
RHS arise from energy conservation at each vertex, and the Kronecker-$\delta$ 
factor excludes non-logarithmic terms which do not alter the impurity state. 
The $\Theta$ step function in Eq. (\ref{eq:RG}) cuts off the 
RG flow of a particular term when the band cutoff $D$ is reduced
below the intermediate-state energy $\Omega_{n\ell}$ of an electron which,
before the scattering process, had the energy $\omega$.
This energy is $\Omega_{n\ell} = \omega+(|n|-|\ell|)\Delta(D)+(\ell-n)B(D)$. 
It depends on the renormalized level spacing $\Delta(D)$ and the 
renormalized Zeeman splitting $B(D)$. The cutoff also involves 
the decay rates of the intermediate impurity states \cite{paaske04b}, 
$\Gamma _m(D)$. Although $\Delta(D)$ and $B(D)$ and $\Gamma_m(D)$, 
contain no leading logarithmic terms, they acquire an RG flow, since they 
are calculated from the 2nd-order self-energy diagrams shown in 
Fig.~\ref{fig:model} c). The 1st-order diagram is real, gives the 
same shift for all impurity states and, hence, does not contribute.
Specifically, $\Delta(D)$ and $B(D)$ are obtained during the RG flow as
\cite{arnold07}, 
\begin{eqnarray}
\hspace{-0.7cm}\Delta(D-\delta D) &=& \Delta(D) - \delta{\rm
Re}\Sigma_0(\nu=0)+\frac{1}{2}\big(\delta{\rm
Re}\Sigma_1(\nu=\Delta+B)+\delta{\rm
Re}\Sigma_{-1}(\nu=\Delta-B)\big)\label{eq:deltaren}\\
\hspace{-0.7cm}B(D-\delta D)&=&B(D)+\frac{1}{2}\big(\delta{\rm Re}\Sigma_1(\nu=\Delta+B)-
\delta{\rm Re}\Sigma_{-1}(\nu=\Delta-B)\big)\ ,
\label{eq:Bren}
\end{eqnarray}
where $\delta{\rm Re}\Sigma={\rm Re}\Sigma(D)-{\rm Re}\Sigma(D-\delta D)$.

\subsection{Two-channel Kondo regime and stability against charge screening}
One of us and collaborators showed in Ref.~\cite{arnold07} that in zero 
magnetic field ($B_0=0$) the model (\ref{eq:hamilton}) has generically a 
stable 2CK fixed point for a wide range of parameters. The 2CK phase occurs, 
because the excited state doublet of the non-interacting impurity is
down-renormalized by Kondo-like interactions below the non-interacting
ground state and the system must then flow to a stable 2CK fixed point.
The Kondo state is formed due to the unbroken SU(2) subgroup of SU(3)
within the (initially excited) doublet of the three-level system. 
Moreover, it was shown that spikes in the differential conductance  
at finite bias voltage arise from Kondo-enhanced transitions between the
ground and the excited states \cite{arnold07}. These spikes are analogous 
to the well-known Kondo satellite resonances observed in 
rare-earth Kondo impurities with several, crystal-field split, local orbitals
\cite{reinert01}. 

It should be emphasized that in the RIM the 2CK Kondo temperature is 
not reduced by charge screening effects: The charge of any impurity is
screened by the conduction electrons. In a TLS in a double-well potential
the spatial charge density distribution is coupled to the presumed 
Kondo pseudospin degree of freedom and is altered by a pseudospinflip
process. Therefore, only those low-energy electrons participate in the Kondo 
scattering which cannot screen the flipping TLS charge distribution  
instantaneously. As shown in Ref.~\cite{aleiner02}, this reduces the energy 
range available for Kondo scattering from the bare conduction bandwidth to 
the TLS tunneling frequency and, thus, suppresses $T_K$. 
By contrast, within the RIM Kondo scattering occurs within 
space-inversion symmetric roton doublets which alters the phase,
but not the charge density distribution of the system. Therefore, charge
density and Kondo degree of freedom are independent, and $T_K$ is not 
influenced by charge screening. 
Moreover, to realize a 2CK regime in the RIM it is not necessary to invoke
transitions via higher excited states in order to enhance $T_K$ over the 
level splitting of the doublet -- which can be prevented by the 
alternating parity of the higher excited states \cite{aleiner01}.
This is because in the RIM the excited-state 
doublet is degenerate by space inversion 
symmetry and because, in any case, a possible level splitting $2B_0$ and 
the Kondo couplings $J_{\perp}$, $J_{z}$ are independent parameters. 
Hence, the 2CK behavior can be cut off only by breaking the
space inversion symmetry, e.g., by lattice distortion, or, in a more 
controlled way, by a magnetic field.

\section{Results in finite magnetic field}
\begin{figure}[t]
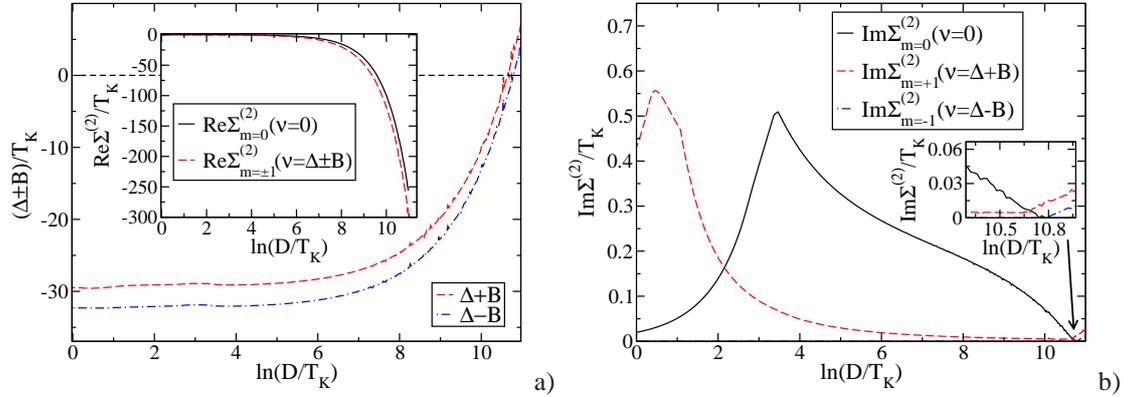

\centering
\includegraphics[scale=0.25]{fig2a.eps}~\,a)
\hfil
\includegraphics[scale=0.25]{fig2b.eps}~\,b)
\caption{\textbf{a)} Perturbative RG flow of the level spacings between the
initially excited $m=\pm1$ states and the $m=0$ state 
for $B_0=1.45\cdot T_K$, 
$T_K=1.7\cdot10^{-5}D_0$, $\Delta_0=5.9\cdot T_K$ and 
$J_{\perp}=J_{||}=0.015\, N(0)$, $g_{m0}^{(j)}=g_{0m}^{(j)}=J_{\perp}/2$. 
$D_0$ is the bare high-energy cutoff. 
The inset shows the RG flow of the real parts of the self-energies for $m=0$ 
and $m=\pm1$.  \textbf{b)} RG flow of the imaginary
part of the 2nd-order impurity level self-energies for the same parameters. 
The inset is a blow-up of the region where the level crossing occurs.}
\label{fig:selfflow}
\end{figure}

We have investigated the appearance of 2CK physics in the three-level 
model (\ref{eq:hamilton}) in the presence of a finite doublet splitting 
or magnetic field, $B_0 > 0$, according to Eqs. (\ref{eq:RG}), 
(\ref{eq:deltaren}) and (\ref{eq:Bren}).
Fig. \ref{fig:selfflow} a) shows an example of the RG flow of the 
excited doublet levels ($m=\pm1$) relative to the initial ground state 
level ($m=0$) for a finite magnetic field $B_0$. $B_0$ was chosen of the order
of the Kondo temperature. $T_K$ was determined here and 
throughout as the cutoff value $D$ for which the dimensionless Kondo coupling 
reaches $N(0)J_\perp(\omega=0)=1$ in zero magnetic field,
with $N(0)$ the density of states at the Fermi level.
The inset of Fig. \ref{fig:selfflow} a) shows the flow of
the on-shell real parts of the impurity self-energies. The differences
$\rm{Re}\Sigma_{m=1}(\Delta+ B)-{\rm Re}\Sigma_{m=0}(0)$ and
$\rm{Re}\Sigma_{m=-1}(\Delta- B)-{\rm Re}\Sigma_{m=0}(0)$ provide the
renormalization of the level spacings $\Delta (D)\pm B (D)$, respectively. 
As the inset shows, this renormalization is negative and 
initially stronger than the bare 
spacings $\Delta_0\pm B_0$, thus causing a level crossing with the $m=0$
state. If this level crossing occurs for both states, $m=\pm1$, as seen in 
Fig.~\ref{fig:selfflow} a), 2CK behavior is realized at the 
lowest energies, involving the two nearly degenerate local levels 
$m=\pm1$ and the two conduction electron channels with magnetic spin 
$s=\pm 1/2$, similar to the $B_0=0$ case \cite{arnold07}. 
Although the perturbative 
RG calculations cannot access this strongly correlated regime, the    
level crossing occurs in general at an early state of the renormalization, 
where the perturbative RG calculations are well controlled, so that the 
occurence of a 2CK fixed point can be safely predicted. 
Towards low energies, the 2CK behavior will be cut off only at the scale 
of the Zeeman splitting $2B(D\to 0)$.  

After the level crossing has occured in the RG flow, the level renormalizations
${\rm Re}\Sigma_{0,\pm1}$ become gradually small, and the level spacings
$\Delta (D)\pm B (D)$ become nearly constant.  
The Zeeman splitting, i.e., the spacing between the $m=+1$ and the 
$m=-1$ states, remains nearly constant during the entire RG flow. 
This level flow can be understood from the behavior of the imaginary parts
of the impurity self-energies, shown in Fig.~\ref{fig:selfflow} b) for the
same set of parameters, since real and imaginary parts are related by 
Kramers-Kronig relations:  
Before the level crossing occures, the two $m=\pm1$ states are excited states 
and, thus, have a large decay rate, proportional to the imaginary parts
of their selfenergies, ${\rm Im}\Sigma_{\pm1}$, while the on-shell decay rate 
of the impurity ground state, $m=0$ is zero (Fig.~\ref{fig:selfflow} b),
inset). Via Kramers-Kronig, this causes a strong, initial down-renormalization 
of the $\Delta (D)\pm B (D)$ and, hence, a level crossing. 
After the level crossing, the $m=-1$ state is the lowest-lying level, 
and its on-shell decay rate vanishes 
[blue, dashed line in Fig.~\ref{fig:selfflow} b)]. The decay rate of the $m=0$ 
state, in turn, starts to grow as the band cutoff $D$ is further reduced, 
due to the increasing coupling constants $g_{0m}^{(j)}(D)$. 
This counter-acts the increase of $\Delta (D)$ and causes it to level-off 
at low energies. Eventually, transitions to the $m=0$ state are frozen out 
when the cutoff is reduced below its (renormalized) excitation energy.  
Below this stage of the renormalization the $m=0$ decay rate decreases 
again [black line in Fig.~\ref{fig:selfflow} b)]. The decay rate of the
Zeeman-split $m=1$ level [red, dashed line in Fig.~\ref{fig:selfflow} b)]
follows that of the $m=0$ level, however shifted towards
lower energies, because of its lower excitation energy, $2B(D)$, 
above the $m=-1$ state. Since for the parameter values of 
Fig.~\ref{fig:selfflow} the initial Zeeman splitting is relatively
small compared to $\Delta$, level renormalizations of the $m=+1$ and the
$m=-1$ state are nearly the same, so that the Zeeman splitting $2B$ remains 
essentially unrenormalized. 
  
\begin{figure}[b]
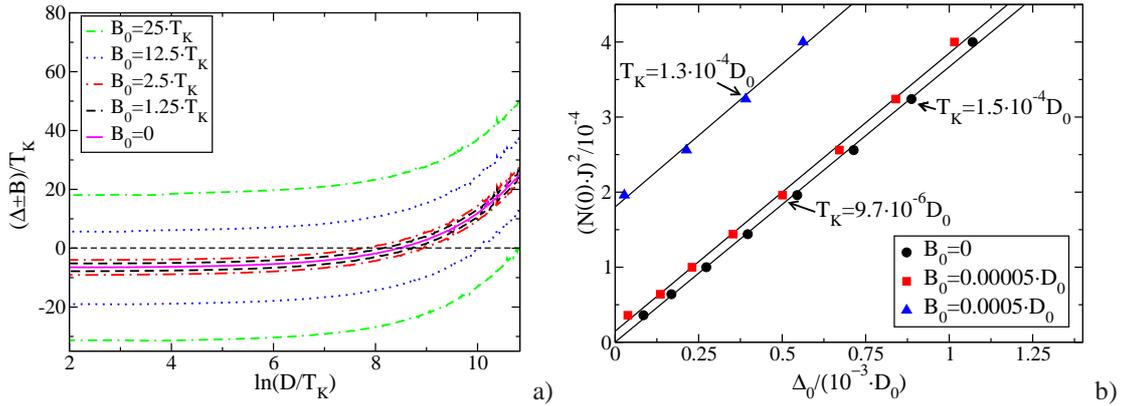

\centering
\includegraphics[scale=0.25]{fig3a.eps}~\,a)
\hfil
\includegraphics[scale=0.25]{fig3b.eps}~\,b)
\caption{\textbf{a)} Level renormalization for
different magnetic fields $B_0$ for $T_K=2\cdot10^{-5}D_0$,
$\Delta_0=25\cdot T_K$ and $J_{\perp}=J_{||}=0.015\,N(0)$,
$g_{m0}^{(j)}=g_{0m}^{(j)}=J_{\perp}/2$. The curves of Zeeman-split doublets
$m=\pm1$ for the same magnetic field $B_0$ are plotted in the same
color. 
\textbf{b)} Phase diagram in the plane of 
$J=J_{\perp}=J_{||}=g_{m0}^{(j)}$ and $\Delta_0$ for
different values of $B_0$. 
The symbols mark the boundary between the 2CK phase (upper left) 
and the potential scattering phase (lower right). The values of $T_K$ 
given in the figure are for the respective parameter values 
$(\Delta_0,J)$ marked by the arrows. 
The solid lines are straight line fits to the data points.}
\label{fig:varB}
\end{figure}

The $m=\pm1$ level renormalization is shown in Fig.~\ref{fig:varB} a) for 
various magnetic fields $B_0$. For stronger magnetic fields the 
level renormalizations, as well as the decay rates 
become slightly smaller, because resonant scattering within the 
$m=\pm1$ doublet is reduced. For magnetic fields up to the order of the 
Kondo temperature $T_K$ and not too large initial level spacing $\Delta_0$, 
both $m=\pm1$ doublet states cross the $m=0$ level. This implies 2CK behavior
in the energy range $B_0 \lesssim E  \lesssim T_K$.
However, if $B_0\gtrsim 10\cdot T_K$, the level crossing occurs only 
for one or none of the $m=\pm1$ states. In this case, the fixed point is a
weak coupling potential scattering impurity.  
These results can be summarized in a phase diagram in terms of the 
coupling constants $J$, $g$ and the level spacings $\Delta_0$ and
$B_0$, see Fig.~\ref{fig:varB} b). The 2CK phase of the system is 
defined as the regime where both $m=\pm1$ levels cross the $m=0$ level.
It occurs to the upper left of the phase boundary lines shown in 
Fig.~\ref{fig:varB} b) for various $B_0$. 
For each value of $B_0$ the critical Kondo coupling $J$ for realizing 
the 2CK phase is in good approximation a quadratic function of the level 
spacing $\delta_0$ [phase boundary lines in Fig.~\ref{fig:varB} b)].
This is a consequence of the fact that for not too large $B_0$ 
the level crossing occurs at
an early stage of the RG flow and, hence, may well be described 
in 2nd-order perturbation theory. Deviations from the quadratic behavior
occur for strong magnetic field [blue triangles in Fig.~\ref{fig:varB} b)],  
where the level crossing occurs later in the RG flow 
[see Fig.~\ref{fig:varB} a)] and 2nd-order perturbation theory is not 
sufficient. For finite $B_0$ the critical Kondo coupling is non-zero even for 
vanishing $\Delta_0$, as expected. However, the 2CK phase is still 
realized in a wide range of parameters.

\section{Discussion and summary}
To conclude, we have proposed and studied a three-level quantum impurity 
model with partially broken SU(3) symmetry, comprised of a degenerate 
excited-state doublet and a unique ground state, which exhibits a generic 
two-channel Kondo (2CK) fixed point. This model may be realized by the 
rotational states of a hydrogen ion (proton) in the interstitial space
of the host lattice of a noble metal. It qualitatively explains two
seemingly distinct features of the differential conductance
experiments by Ralph and Buhrman \cite{ralph92,ralph95} on the same footing, 
namely the zero-bias anomaly (ZBA) characteristic of 2CK 
behavior and the conductance spikes at elevated bias. The 
ZBA is due to a down-renormalization of the doublet below the 
non-interacting impurity ground state and Kondo scattering from this 
doublet, while the spikes at elevated bias are due to Kondo-enhanced 
transitions between impurity ground and excited states \cite{arnold07}.
We have discussed that charge screening effects do not reduce the Kondo 
temperature $T_K$ of the present rotational impurity model,
in contrast to the case of two-level systems \cite{aleiner02}.
Since Kondo coupling and splitting of the doublet are independently 
controllable parameters of the model, Kondo transitions via virtual
excitations of higher excited states are not required in order to 
enhance the Kondo tmeperature. Rather, real transitions to excited states,
induced by finite bias, are expected to lead to multiple conductance spikes
\cite{arnold07}, which are also observed experimentally \cite{ralph92,ralph95}. 
We have also analyzed the effect of a Zeeman splitting of the rotational
doublet by an external magnetic field $B$. A moderately strong field
$B < T_K$ does not destroy the 2CK phase of the model, although 
the 2CK zero-bias conducance anomaly will be cut off at the lowest 
energies by the Zeeman splitting. This behavior of the model
is also observed experimentally \cite{ralph95}. 
In the three-level model the Zeeman splitting of the 
$m=\pm1$ doublet also splits the transition energies $\Delta\pm B$ 
between the $m=0$ and the $m=+1$ or $m=-1$ states, respectively,
which mark the positions of the finite-bias conductance spikes. 
A splitting of the conductance spikes in a magnetic field is, 
therefore, expected if their width is smaller than the Zeeman energy. 
This will be investigated in forthcoming work \cite{ballmann12}.

We are indepted to L. Borda, E. Fuh Chuo and especially A. Zawadowski 
for numerous useful discussions. 
On the occasion of his 60th birthday, we dedicate this article to 
Ulrich Eckern, who has been a valued colleague and good friend for so 
many years. This work was supported by DFG through SFB 608 and 
grant No. KR1762/2. 

%

\begin{thebibliography}{[1]}

\bibitem{nozieres80}
P.~Nozi\`{e}res and A.~Blandin,
Journal de Physique (Paris), {\bf 41}, 193 (1980).

\bibitem{kolf07}
Ch.~Kolf and J.~Kroha, 
Phys.~Rev.~B~{\bf 75}, 045129 (2007).

\bibitem{andrei84}
N.~Andrei and C.~Destri,
Phys.~Rev.~Lett. {\bf 52}, 364 (1984).

\bibitem{tsvelik94}
A.~M.~Tsvelik and P.~B.~Wiegmann,
Z.~Phys.~B {\bf 54}, 201 (1994).

\bibitem{seaman91}
C. L. Seaman, M. B. Maple, B. W. Lee, S. Ghamaty,
M. S. Torikachvili, J.-S. Kang, L. Z. Liu, J. W. Allen, and D. L. Cox, 
Phys.~Rev.~Lett. {\bf 67}, 2882 (1991);

\bibitem{cichorek05}
T. Cichorek, A. Sanchez, P. Gegenwart, F. Weickert, A. Wojakowski, 
Z. Henkie, G. Auffermann, S. Paschen, R. Kniep, and F. Steglich,
Phys.~Rev.~Lett.~{\bf 94}, 236603 (2005).

\bibitem{cox88a}
D. L. Cox, Phys. Rev. Lett. {\bf 59}, 1240 (1987).

\bibitem{cox88b}
D. L. Cox, J. Mag. Mag. Mat.  {\bf 76}, 53 (1988).

\bibitem{oreg03}
Y.~Oreg and D.~Goldhaber-Gordon, 
Phys.~Rev.~Lett.~{\bf 90}, 136602 (2003).

\bibitem{goldhaber07}
R.~M.~Potok, I.~G.~Rau, H.~Shtrikman, Y.~Oreg, and D.~Goldhaber-Gordon,
Nature {\bf 446}, 167 (2007).

\bibitem{ralph92}
D.~C.~Ralph and R.~A.~Buhrman,
Phys.~Rev.~Lett. {\bf 69}, 2118 (1992);

\bibitem{ralph95}
D.~C.~Ralph and R.~A.~Buhrman,
Phys.~Rev.~B {\bf 51} 3554 (1995).

\bibitem{affleck91}
I.~Affleck and A.~W.~W.~Ludwig,
Nucl.~Phys.~B, {\bf 352}, 849 (1991).

\bibitem{affleck93}
I.~Affleck and A.~W.~W.~Ludwig,
Phys.~Rev.~B~{\bf 48}, 7297 (1993).

\bibitem{vdelft95}
D.~C.~Ralph, A.~W.~W.~Ludwig, J. von Delft, and R.~A.~Buhrman,
Phys.~Rev.~Lett.~{\bf 75}, 770 (1995).

\bibitem{altshuler79}
B. L. Altshuler and A. G. Aronov, Sov. Phys. JETP {\bf 50}, 968 (1979).

\bibitem{kozub97}
V.I.~Kozub~and~A.M.~Rudin,~
Phys.~Rev.~B~{\bf 55},~259~(1997).

\bibitem{hettler94}
M.~H.~Hettler, J.~Kroha, and S.~Hershfield,
Phys.~Rev.~Lett.~{\bf 73}, 1967 (1994).

\bibitem{vondelft98}
J. von Delft, D. C. Ralph, R. A.  Buhrman, S. K. Upadhyay, 
R. N. Louie, A. W. W. Ludwig, and V. Ambegaokar,
Ann. Physics (New York) {\bf 263}, 1 (1998).

\bibitem{coxzawa98}
D.~L.~Cox and A.~Zawadowski,
Adv.~Phys. {\bf 47}, 599 (1998).

\bibitem{vladar83}
K.~Vlad\'{a}r and A.~Zawadowski,
Phys.~Rev.~B {\bf 28}, 1564 (1983); {\bf 28}, 1582 (1983); 
Phys.~Rev.~B {\bf 28}, 1596 (1983). 

\bibitem{vladar88}
K.~Vlad\'{a}r, A.~Zawadowski, and G. T. Zimanyi,
Phys.~Rev.~B {\bf 37}, 2015 (1988).

\bibitem{zawa94}
G. Zar\'and and A. Zawadowski, 
Phys.~Rev.~Lett.~{\bf 72}, 542 (1994);
Phys.~Rev.~B~{\bf 50}, 932 (1994).

\bibitem{aleiner02}
I.~L.~Aleiner and D.~Controzzi,
Phys.~Rev.~B {\bf 66}, 045107 (2002).

\bibitem{aleiner01}
I.~L.~Aleiner, B. L. Altshuler, Y. M. Galperin, and T. A. Shutenko,
Phys.~Rev.~Lett.~{\bf 86}, 2629 (2001).
 
\bibitem{zarand05}
G. Zar\'{a}nd,
Phys.~Rev.~B {\bf 72}, 245103 (2005).

\bibitem{arnold07}
 M. Arnold, T. Langenbruch, and J. Kroha, PRL \textbf {99}, 186601 (2007).

\bibitem{rosch03}
A.~Rosch, J.~Paaske, J.~Kroha, and P.~W\"olfle,
Phys.~Rev.~Lett.~{\bf 90}, 076804 (2003).

\bibitem{paaske04a}
J.~Paaske, A.~Rosch, and P.~W\"olfle,
Phys.~Rev.~B~{\bf 69}, 155330 (2004).

\bibitem{rosch05}
A. Rosch, J. Paaske, J. Kroha, and P. W\"olfle, 
J. Phys. Soc. Jpn. {\bf 74}, 118 (2005).


\bibitem{paaske04b}
J.~Paaske, A.~Rosch, J.~Kroha, and P.~W\"olfle,
Phys.~Rev.~B~{\bf 70}, 155301 (2004).

\bibitem{reinert01}
F. Reinert {\it et al.}, Phys. Rev. Lett. {\bf 87}, 106401 (2001);
D. Ehm {\it et al.}, Phys. Rev. B {\bf 76}, 045117 (2007).

\bibitem{ballmann12}
K. Ballmann and J. Kroha, in preparation.
\end{thebibliography}
%

\end{document}